\documentstyle[12pt]{article}
\newcommand{\be}{\begin{equation}}

\newcommand{\ee}{\end{equation}}
\newcommand{\bea}{\begin{eqnarray}}
\newcommand{\eea}{\end{eqnarray}}

\newcommand{\bel}[1]{\be\label{#1}}
\newcommand{\re}[1]{Eq.~(\ref{#1})}
\newcommand{\mbs}[1]{\mbox{$\scriptstyle{#1}$}}
\newcommand{\ds}{\displaystyle}

\newcommand{\loo}{\,\raisebox{-.5ex}{$\stackrel{<}{\scriptstyle\sim}$}\,}

\begin{document}

\baselineskip 24pt
\begin{center}
%---------------------------------------------------------
{\Large\bf Collective mechanism of dilepton\\
production in high--energy nuclear collisions}\\[5mm]
%---------------------------------------------------------
{\bf I.N.~Mishustin$^{a,b}$, L.M.~Satarov$^a$, H.~St\"ocker$^c$ and
W.~Greiner$^c$}
\end{center}
\baselineskip 16pt
\begin{tabbing}
\hspace*{4em}\=${}^a$\,\={\it The Kurchatov~Institute,
123182~Moscow,~\mbox{Russia}}\\
\>${}^b$\>{\it The Niels~Bohr~Institute,
DK--2100~Copenhagen {\O},~\mbox{Denmark}}\\
\>${}^c$\>{\it Institut~f\"{u}r~Theoretische~Physik,
J.W.~Goethe~Universit\"{a}t,}\\
\>\>{\it D--60054~Frankfurt~am~Main,~\mbox{Germany}}
\end{tabbing}
\baselineskip 24pt

\begin{abstract}
Collective bremsstrahlung of vector meson fields
in relativistic nuclear collisions 
is studied within the time--dependent Walecka model.
Mutual deceleration of the colliding nuclei is described by
introducing the effective stopping time and average rapidity 
loss of baryons. It is shown that electromagnetic 
decays of virtual $\omega$--mesons produced by bremsstrahlung
mechanism can provide a substantial contribution to the soft dilepton
yield at the SPS bombarding energies. In particular, it may be 
responsible for the dilepton enhancement observed in 160 AGev
central Pb+Au collisions. Suggestions for future experiments
to estimate the relative contribution of the collective mechanism 
are given.   
\end{abstract}

A high degree of baryon stopping observed in relativistic 
nuclear collisions implies a strong space--time variation
of the collective baryon current in these reactions.
According to the time--dependent Walecka model~\cite{Wal85},
in this situation it is natural to expect the excitation
of propagating waves (bremsstrahlung) of classical meson
fields. Conversion of virtual mesons into real particles may provide
an additional important source of secondary particles 
in nuclear collisions. Unlike the conventional mechanism 
of incoherent hadron--hadron collisions the contribution
of this source increases strongly (quadratically) with the number
of participating nucleons. Following this idea in Refs.~[2-4] 
we considered the production of pions, baryon--antibaryon 
pairs and dileptons by the collective bremsstrahlung 
and decay of virtual $\omega$--mesons. 
It was shown that the role of the Collective Bremsstrahlung
Mechanism (CBM) grows with bombarding energy and that it  
becomes important in central collisions 
of heavy nuclei already at the SPS energies.

The study of dilepton production is especially interesting
because of the strong enhancement of soft dileptons observed
in nuclear collisions by the CERES~\cite{Cer95} and 
HELIOS--3~\cite{Hel95} collaborations. It was 
shown~[7--11] that the contribution  of $\pi\pi\to\rho\to l^+l^-$
processes may explain the enhancement only if one assumes  
a strong modification of $\rho$--mesons at intermediate
stages of a heavy--ion collision. On the other hand,
recent calculations of the $\rho$--meson spectral function 
show~\cite{Ra97,Cas97} that the $\rho$ peak exhibits 
a substantial broadening in nuclear matter. In this case it
is questionable if the $\rho$--meson can  
be treated as a well--defined quasiparticle.
Below we show that the observed enhancement of the dilepton yield
may be explained, at least partly, by the contribution of the CBM. 

The detailed formulation of the dilepton production within the CBM 
is described in Ref.~\cite{Mis97}. Below we outline only the main 
points of the model. The 4--momentum distribution of primordial 
$\omega$--mesons produced by the bremsstrahlung mechanism
is determined by the Fourier transform of the collective
baryonic current $J_\mu (p)$:
\bel{dvom}
\frac{\ds{\rm d}^4 N_{\omega}}
{\ds{\rm d}^4p\hfill}= \frac{g_V^2}{16\pi^3}|J_\mu^*(p)J^\mu(p)|
\rho_\omega(p^2)\,,
\ee 
where $\rho_\omega(p^2)$ is the spectral function of virtual $\omega$'s
describing their deviation from the mass shell 
$p^2=m_\omega^2$\,. This spectral function is written in the 
Breit--Wigner form~\cite{Mis97} with the mass $m_\omega$
and width $\Gamma_\omega$.   

The mass distribution of dileptons produced by the CBM
can be written as a convolution
of the virtual $\omega$ meson spectrum and the differential branching
of the \mbox{$\omega\to l^+l^-X$} decay ($X$~denotes any particle
emitted together with the lepton pair):
\bel{mdil}
\frac{\ds{\rm d}N_{l^+l^-}}{\ds{\rm d}M\hfill}=\int{\rm d}^4p
\frac{\ds{\rm d}^4N_{\omega}}{\ds{\rm d}^4p}\cdot
\frac{\ds{\rm d}B_{\mbs{\omega\to l^+l^-X}}}{\ds{\rm d}M\hfill}\,,
\ee
where $M$ is the dilepton invariant mass. The procedure
of calculating the partial widths of the off--mass-shell 
$\omega$--mesons is described in Ref.~\cite{Mis97}. 
In our calculations we take into account the direct 
($\omega\to l^+l^-$) as well as the Dalitz ($\omega\to\pi^0l^+l^-$) 
decay channels.

According to \re{dvom} the collective bremsstrahlung may be 
important if the Fourier--trans\-formed baryonic current is nonzero
in the time--like region $p^2\sim M^2$, i.e. if 
the projectile--target mutual deceleration is strong
enough. Simple estimates show~\cite{Mis95} that noticeable effects
may be expected if the effective deceleration time $\tau_d$ is
shorter than $\hbar/M$\,. 
To calculate $J^\mu(p)$ we adopt the schematic picture
of a heavy--ion collision suggested in Ref.~\cite{Vas80}.   
We consider central collisions of equal nuclei in the cm frame
and assume that $\tau_d$ is equal to the passage time, 
$\tau_d=2R/\sinh{y_0}$, where $R$ is the geometrical
radius of the colliding nuclei and $y_0$ is the initial cm 
rapidity. In our dynamical picture the effects of internal compression 
and transverse motion are disregarded. The interpenetrating nuclei 
move as a whole along the beam axis with instantaneous 
velocities $\pm\dot{z}(t)$ which are parametrized as 
\bel{prtr}
\dot{z}(t)=v_f+\frac{\ds v_0-v_f}{\ds 1+{\rm e}^{2t/\tau_d}}\,.
\ee
Here $v_0=\tanh{y_0}$ and $v_f$ is the final velocity of nuclei
at $t\to +\infty$. The degree of baryonic stopping 
is characterized by the average rapidity
loss $\delta y=y_0-{\rm Artanh} v_f$. For central
Au+Au collisions at SPS energies we take the value $\delta y$=2.4 
motivated by RQMD calculations~\cite{Mis95}. The time integrals 
determining $J^\mu(p)$ within this dynamical model are calculated 
numerically~\cite{Mis97} assuming the Woods--Saxon density distribution 
in the colliding nuclei. In all calculations we take $g_V=13.78$.

The resulting mass spectrum of $e^+e^-$ pairs produced in central
160 AGeV Au+Au collisions is shown in Fig.~1. The calculations
were made assuming that the on--mass--shell \mbox{$\omega$--mesons} 
have the vacuum mass, $m_\omega=0.782$ GeV and decay
width, \mbox{$\Gamma_\omega=\Gamma_0=8.4$ MeV}.  
We take into account the same
acceptance cuts as used in the CERES measurements of the dilepton
yield in Pb+Au collisions~\cite{Cer95}. The jump
of the predicted spectrum at $M=0.35$ GeV is caused by the cut at
low transverse momenta of leptons. Note that the mass resolution 
corrections are not included in this calculation. One can see
that the CBM contribution is important in the same region
of intermediate masses, \mbox{$M=0.35-0.8$ GeV}, where the enhanced
dilepton production is observed experimentally.

Of course, dileptons are also produced by the
incoherent formation and decay of pions 
and hadronic resonances. This hadronic ''cocktail''
contribution has been obtained~\cite{Cer95} 
by the Monte Carlo simulation which includes 
decays of pions and 
known meson resonances. The relative abundances of 
resonances have been found by scaling the pp and pA data.
According to Fig.~1, the incoherent dilepton source   
is more important outside the intermediate 
mass region. The sum of the CBM and cocktail contributions 
satisfactorily reproduces the CERES data. However, the data are 
still slightly underestimated at \mbox{$M\simeq 0.5-0.6$ GeV} and 
a too sharp peak is predicted at $M\simeq m_\omega$\,.

As shown in Refs.~\cite{Wo97,Kli97} the $\omega$--meson width may
increase significantly in dense baryonic matter. At baryon
densities $n\sim n_0$, where $n_0$ is the normal nuclear density,
the in--medium width of $\omega$--mesons 
may exceed $\Gamma_0$ by an order 
of magnitude. To estimate the effects of the in--medium broadening of
virtual $\omega$--mesons, we performed the same calculation as above, but
replacing $\Gamma_0$ by a higher width $\Gamma_\omega=5\Gamma_0$\,. 
Fig.~2 shows that the agreement with the data becomes much
better in this case.

Spectra of primordial $\omega$'s produced by the CBM have two 
characteristic features~\cite{Mis95}. First their $p_T$ distributions
are relatively soft due to the cutoff $p_T\loo\hbar/R$
imposed by the nuclear form factor. Second,
the rapidity spectra of these $\omega$--mesons
have two side maxima, symmetrical with respect to $y_{\rm c.m.}=$0\,.
Hopefully, these features can be seen in the distributions over
the total momentum of dileptons. Such measurements   
can be used in future experiments to extract the collective component
above the ''background'' of the incoherent production. 

To illustrate this model prediction we calculated 
the distributions over
the total rapidity of a dilepton pair produced by the CBM 
in the central Au+Au collision at 160 AGeV.
The results are shown in fig.~3. Unlike the preceding
figures, no acceptance cuts are included in this calculation.
A strong mass dependence of the dilepton yield, 
more pronounced at small 
total rapidities in the cm frame, is clearly seen. Also one can see 
the dip at $y_{c.m.}=0$
which is the characteristic signature of the CBM~\cite{Mis95}.
The dip is partly filled up due to presence of Dalitz
decays. Nevertheless, it is well seen, especially at higher 
masses where the Dalitz contribution is relatively small. 
     
In conclusion, we have shown that the CBM may provide an important
source of dileptons in high--energy nuclear collisions.
This mechanism can be responsible, at least partly,
for the enhanced yield of dileptons observed at the SPS bombarding 
energies. In order to observe the collective bremsstrahlung contribution
it is highly desirable to extend the experimental acceptance
to lower $p_T$, to improve the mass resolution and to measure 
the rapidity dependence of the dilepton yield.

The authors thank for the support from Graduirtenkolleg (Universit\"at
Frankfurt), DFG, BMFT and the Carlsberg Foundation.

%------------------------------------------------------
\section *{Figure captions}
%------------------------------------------------------
\newcommand{\Fig}[2]{\noindent{\bf Fig.~#1.~}
\parbox[t]{15cm}{\baselineskip 24pt #2}\\[5mm]}
\Fig{1}
{Invariant mass distributions of $e^+e^-$ pairs in central 160 AGeV Au+Au
collisions. The collective bremsstrahlung contribution
is shown by the dashed line. The dotted line is  
the result of Monte--Carlo simulation (hadronic cocktail) from
Ref.~\cite{Cer95}.
The solid curve gives the sum of these two mechanisms.
Preliminary experimental data for central Pb+Au collisions
are taken from Ref.~\cite{Cer95}.}
\Fig{2}
{The same as Fig. 1, but with the dashed line calculated for  
the increased $\omega$--meson width 
$\Gamma_\omega=5\,\Gamma_0$=42 MeV.}
\Fig{3}
{Total rapidity distribution of $e^+e^-$ pairs produced by 
bremsstrahlung mechanism at different invariant masses $M$\,.
No acceptance cuts are included in the calculation.}
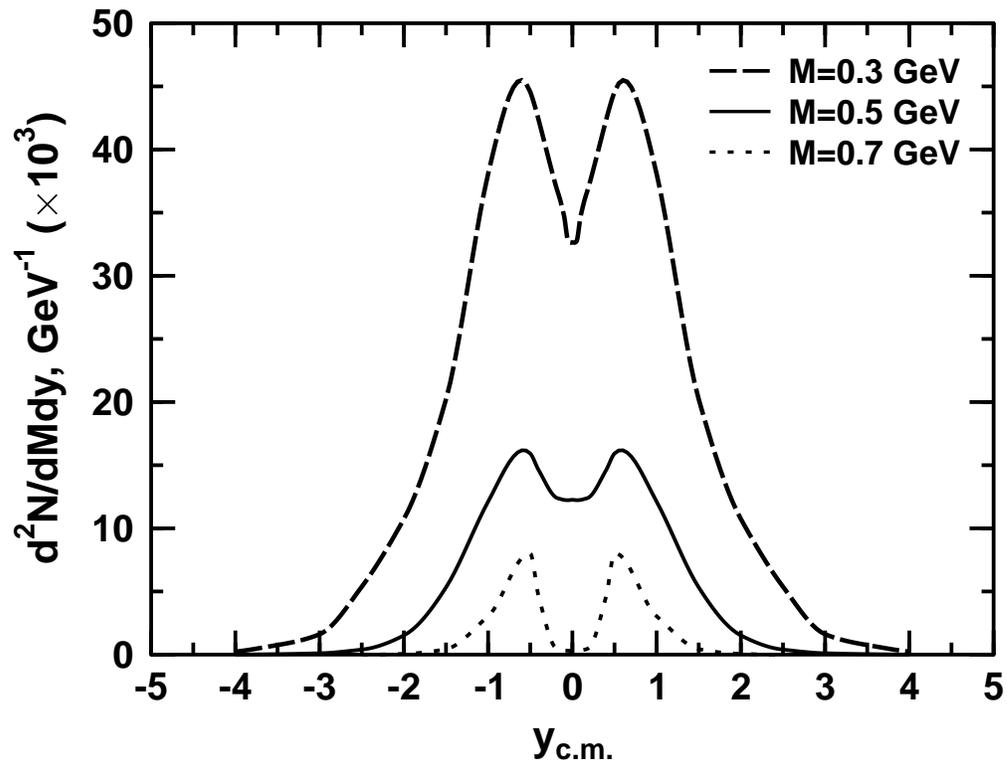
\end{document}